\documentclass[aps,prl,twocolumn,groupedaddress]{revtex4}
\usepackage{graphicx}
\begin{document}


\title{Effects of rf Current on Spin Transfer Torque Induced Dynamics}


\author{S. H. Florez}
\email[]{sylvia.florez@hitachigst.com}
\author{J. A. Katine}
\author{M. Carey}
\author{L. Folks}
\author{O. Ozatay}
\author{B. D. Terris}

\affiliation{Hitachi Global Storage Technologies, San Jose Research Center, 3403 Yerba Buena Road, San Jose CA 95135}


\date{\today}

\begin{abstract}
The impact of radiofrequency (rf) currents on the direct current (dc) driven switching dynamics in current-perpendicular-to-plane nanoscale spin valves is demonstrated. The rf currents dramatically alter the dc driven free layer magnetization reversal dynamics as well as the dc switching level. This occurs when the frequency of the rf current is tuned to a frequency range around the dc driven magnetization precession frequencies.  For these frequencies, interactions between the dc driven precession and the injected rf induce frequency locking and frequency pulling effects that lead to a measurable dependence of the critical switching current on the frequency of the injected rf. Based on macrospin simulations, including dc as well as rf spin torque currents, we explain the origin of the observed effects.
\end{abstract}

\maketitle
\section{INTRODUCTION}
Direct electrical currents, via the spin transfer torque (STT), can excite a broad variety of magnetization dynamics, including reversal and steady-state precessional states in current-perpendicular-to-plane spin valves. These precessional dynamics, stabilized by the opposing effects of the STT and the damping torque, in addition to being of great scientific interest have important implications for the development of new radiofrequency (rf) devices  \cite{Berger,Slonzcewski}. While prior studies focused on the dynamics in the non-hysteretic regime, it has also been recognized that precessional dynamics can exist in the hysteretic regime \cite{Kiselev,Devolder,Sun,Stiles}. These pre-switching (PS) precessional modes are of great interest as they are responsible for driving the switching dynamics and control of switching is key to potential memory and logic applications.

In this work we investigate the highly unexplored PS dynamical regime as well as the effects of additional rf currents on the PS and switching modes. The PS modes are found to interact with the applied rf currents having frequencies in the range of the dc-only driven modes, and under certain conditions are found to frequency lock. In contrast to previous frequency locking results \cite{Rippard, Tulapurkar, Sankey}, by looking at these effects in the PS regime, we find a measurable dependence of the dc switching threshold on the external rf frequency, $f_{inj}$.  In the locking range the rf appears to have a stabilizing effect which hinders switching, however, at frequencies just below this range, we find a measurable reduction in the direct critical current for switching, $I_{c}$. While we have already reported this effect \cite{Florez}, an explanation for the origin of this effect has not been given. In this work, by studying the behavior of this system in frequency domain and through macrospin simulations (MS) (using the Landau-Lifshitz-Gilbert equation with Slonczewski STT \cite{Slonzcewski}) we provide a description of the magnetization dynamics originating from the interaction between the dc driven switching modes and the applied rf. The results show good agreement with our experimental data.

\section{EXPERIMENT}
The samples are 50 nm x 100 nm hexagonal pillars comprised of an IrMn pinned antiparallel (AP) coupled bilayer $Co_{50}Fe_{50}(25{\AA})/Ru(8{\AA})/Co_{50}Fe_{50}(25{\AA})$, a 40 ${\AA}$ Cu spacer, and 35 ${\AA}$ $Ni_{92}Fe_{8}$ as the free layer (FL). We assume all observed current induced magnetic excitations occur in the FL. In our system of reference, positive fields/negative currents stabilize the AP magnetization state. Here we present results from a single element, selected due to its comparatively high power PS modes even in absence of applied external field $H_{app}$.  This is likely due to the significant dipolar field $H_{dip}$  $\sim$ 170 Oe, at the free layer (from the field hysteresis curve offset), despite the use of AP coupled pinning layers.
We point out, however, that variations from one element to another are observed and furthermore, PS dynamics are not detectable for all pillars. Nevertheless, qualitatively similar results were obtained for all samples for which similar amplitude PS modes were observed.

Experiments were performed in a variable temperature \textit{Desert Cryogenics} probe station equipped with a superconducting magnet for applying in-plane magnetic fields.  Contact to the pillar was made through high frequency ($<$40GHz) probes connected to the (rf+dc) mixing port of a broadband bias tee. The inductive port of the bias tee was used for low frequency dynamic resistance measurements using a lock-in amplifier, while the capacitive side was attached to a splitter that reroutes rf signals from a high frequency signal generator and toward a + 44 dB amplifier attached to a spectrum analyzer.  While the sample stage was held at 4.2 K, the true sample temperature, when injecting direct currents, $I^{dc}$ on the order of 1 mA, was estimated to be approximately 25 K due to Joule heating. All experiments were performed at low temperatures in order to minimize thermal fluctuations.
Calibration of the electrical set-up was carried out by measuring the temperature dependent Johnson noise from 50 Ohm test resistors, and a network analyzer was used to measure the complex and frequency dependent impedance $Z_{0}$ of the sample, which was then used to estimate the rms amplitude of the injected rf current.
\section{RESULTS AND DISCUSSION}
The macrospin model predicts the onset of sub-critical current driven stable precessional dynamics around the local dipolar field for currents above a threshold value $I_{t}$, for which the STT balances the damping torque. As $I^{dc}$ increases, the orbits are predicted to expand while precession frequencies decrease \cite{Sun, Stiles}. Experimentally, we identify potential regions of STT induced dynamic instabilities in the FL magnetization as anomalies in $dV/dI$.   Fig.$\:$1(a) illustrates this for current sweeps taken at $H_{app}$ =25, 50 and 100 Oe, where the field is applied a few degrees off the easy axis direction. We concentrate mainly on the STT magnetization dynamics prior to the AP to P transition, occurring approximately at -560 Oe at $I^{dc}$=0 (not shown) and 1.45 mA at $H_{app}$=0.

At currents greater than the instability threshold current $I_{t}$, anomalies, appearing as small dips for $H_{app}$=25, 50 Oe (hysteretic) and larger dips for $H_{app}$=100 Oe (non-hysteretic) in Fig.$\:$1(a), reveal strong microwave activity. Shown in Fig.$\:$1(b) is a phase diagram, where the color scale represents $dV/dI$ for $I^{dc}$ swept from negative to positive values and the continuous lines show the switching boundaries obtained for current sweeps in both directions.  The PS regime appears as a narrow bright area, on the right side of the dotted line and to the left of the P region. Low positive fields ($<$100 Oe), that oppose STT yet sustain hysteretic behavior, stabilize and strengthen these precessional states over a wider $I^{dc}$ range, as can be seen through the widening of the PS regime. Fig.$\:$ 1(c) compares current driven dV/dI loops measured with and without injected rf at $H_{app}=0$,  showing the strong anomalies induced by the rf.  We have observed this type of behavior for samples with two distinct FL materials \cite{Florez}.
\begin{figure}
\includegraphics[width=6.5 cm]{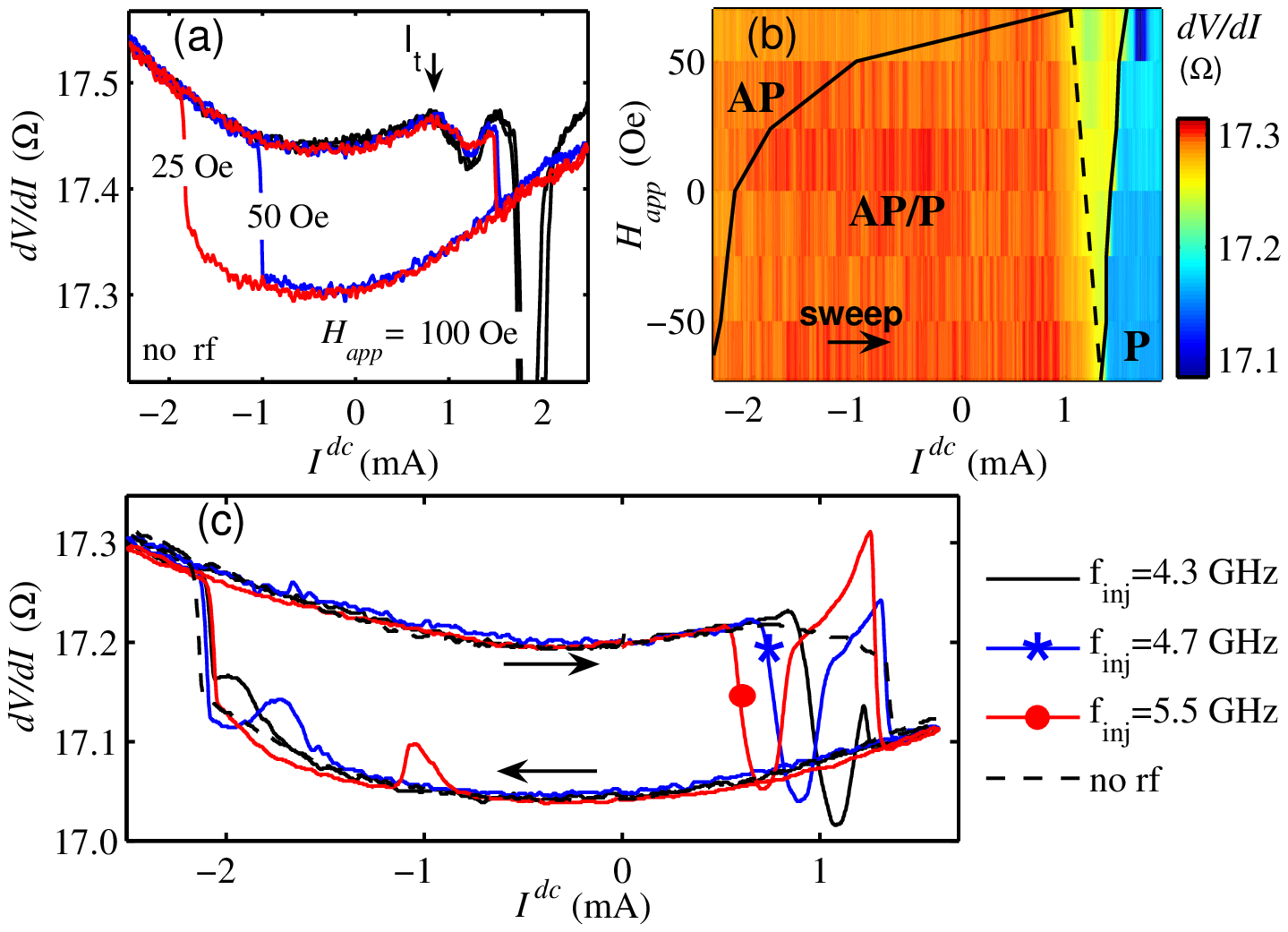}
\caption{(Color online)(a) $dV/dI$ vs. current induced switching loops for $H_{app}$= 25, 50 and 100 Oe, with $I_{t}$ marked by arrow. (b) ($I^{dc}$, $H_{app}$) phase diagram for hysteretic region with color scale representing $dV/dI$. (c) $dV/dI$ vs. current induced switching loops for $H_{app}$= 0 with no rf (dashed) and injected rf (continuous) with $f_{inj}$=4.3, 4.7 and 5.5 GHz.}
\end{figure}

Examples of power spectral density (PSD) data in the PS regime are shown in Fig.$\:$ 2(a). As $I^{dc}$ increases in this range, the emitted power rises as the dc driven resonance frequency $f_{0}^{dc}$ drops (\emph{redshifts}) in agreement with the macrospin model. In Fig.$\:$ 2(b) we plot the measured $f_{0}^{dc}$ vs $I^{dc}$ for $H_{app}$=0 and 50 Oe, showing the peak frequencies decrease as the current rises toward the switching value $I_{c}$.  Extrapolation of these data through $I_{c}$ suggests that the minimum resonance frequencies approaching reversal $(f_{0}^{dc})^{min}$ and for $H_{app}$=0 are near $4.2$ GHz. Following Kiselev et al. \cite{Kiselev}, we make a macrospin estimation of the precession and misalignment angles (between the pinned layer and the FL) by assuming small angle precession and measuring the total integrated power in the first and second harmonic.  While this approximation may be imprecise, in view of the highly non-Lorenztian spectra, it provides a rough picture of the behavior of the precessional dynamics as $I^{dc}$ is ramped. We find precession angles increasing from 10 to 30 degrees, for $I^{dc}$ increasing from 1 to 1.4 mA and for 0 $<$ $H_{app}$ $<$ 100 Oe. However, the multiple peak structure and broad linewidths, reaching up to $\sim$ 500 MHz, suggest incoherence and coexistence of several modes. Such incoherent behavior has been observed in micromagnetic modeling \cite{Lee}, and can be understood from the spatially non-uniform local demagnetizing fields at the edges of the spin-valve.

As previously stated, the effect of adding an rf current, $I^{rf}$ to $I^{dc}$ on the FL STT dynamics can be seen from the $dV/dI$ vs current hysteresis loops in Fig.$\:$1(c).  This figure includes data for the no rf case (dashed line) and for three different injected frequencies $f_{inj}$ =4.3, 4.7 and 5.5 GHz. In all cases, a power of -15 dBm was applied at the signal generator, resulting in $(I^{rf})^{rms} \sim 1 mA$.  The dips (or peaks) appearing before the AP to P (or P to AP) transition are caused by interactions between $I^{rf}$ and the $I^{dc}$ driven precession resulting in changes in the FL magnetization precessional trajectory and in the magnetoresistance. Focusing on the AP to P switch, note that as $f_{inj}$ increases, the $I^{dc}$ level at which the dip occurs decreases, following the $f_{0}^{dc}$ increase induced by decreasing $I^{dc}$. As discussed below, for external frequencies below the $f_{0}^{dc}$ range, we observe a reduction in $I_{c}$.
\begin{figure}
\includegraphics[width=7 cm]{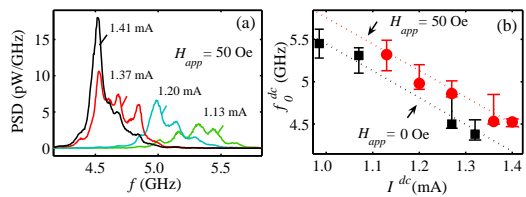}
\caption{(Color online)(a) Power spectral density measured at $H_{app}$=50 for $I^{dc}$  values below the switching value. The PSD measured with $I^{dc}$=0 has been subtracted. (b) $f_{0}^{dc}$ vs. $I^{dc}$ for $H_{app}$=0 (squares), 50 Oe (circles) (symbols at peak value, bar denotes FWHM span).}
\end{figure}

To elucidate the effects of the rf on the FL dynamics, we studied the changes in the frequency domain behavior induced by adding $I^{rf}$ to $I^{dc}$. Results suggest partial frequency locking between the dc driven precession and the external rf signal as shown in Fig.$\:$3(a).  For these experiments power was limited to -23 dBm to avoid saturating the amplifier.  Low applied fields below 50 Oe were applied to enhance the signal originating at the sample.   Fig.$\:$3(a) illustrates the partial frequency locking of the applied and internal oscillations by comparing signals obtained at $I^{dc}$=1.40 and 1.44 mA, with (after subtracting the external rf signal) and without external rf. While the peak frequency for $I^{rf}$=0 is at 4.8 GHz, applying $I^{rf}$ with $f_{inj}$ at 4.5 GHz or 4.75 GHz narrows the signal and shifts the measured power to the $f_{inj}$ value. As expected, locking only occurs when $I^{rf}$ is applied at a frequency at which the dc driven mode also has significant power. Further measurements (not shown here) suggest a correlation between the existence of dips/peaks in $dV/dI$ and the ($I^{dc}$, $f_{inj}$) values for which frequency locking occurs.

\begin{figure}
\includegraphics[width=6.5 cm]{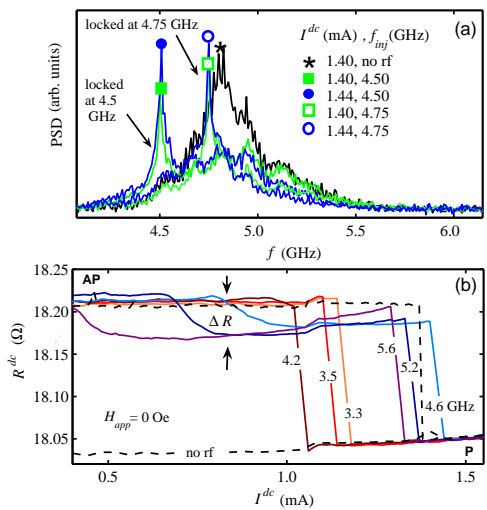}
\caption{(Color online)(a) Comparison of the PSD with (circles, squares) and without (star) rf for $I^{dc}$ in the PS range and $H_{app}$=25 Oe, showing frequency locking for $f_{inj}$  =4.5 GHz and 4.75 GHz. (b) $R^{dc}$  for $I^{dc}$ near the AP to P switch, and for several values of injected frequencies. The dashed line shows the no rf case.}
\end{figure}
Additional insight can be extracted from dc resistance ($R^{dc}$) measurements. As shown in Fig.$\:$3(b), for $f_{inj}$ close to $f_{0}^{dc}$, a reversible plateau forms prior to switching, with a $\Delta R$ of approximately $ 0.05$ $\Omega $. The effect appears at lower values of dc for higher injected frequencies. Our data confirms that the ($I^{dc}$, $f_{inj}$) values for which the dips/peaks in $dV/dI$ form and for which frequency locking occurs also coincide with the formation of the plateau in the dc resistance. In accordance with the previously observed behavior, the $R_{dc}$ data shows that for $f_{inj}$ in the range $(f_{0}^{dc})^{min} \lesssim 4.2 $GHz, the plateau does not form and $I_{c}$ drops as the $f_{inj}$ increases.

\begin{figure}
\includegraphics[width=4.5 cm]{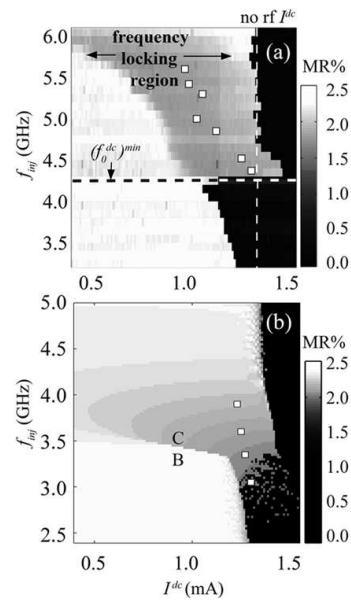}
\caption{(a)($I^{dc}$, $f_{inj}$) phase map obtained experimentally at $H_{app}=0 $ with $R^{dc}$ represented through gray scale. The white squares show the locations of the measured $f_{0}^{dc}$. (b) ($I^{dc}$, $f_{inj}$) phase map obtained through MS, with gray scale representing average final state MR. The white squares show the locations of the $f_{0}^{dc}$ obtained numerically.}
\end{figure}

In Fig.$\:$4(a) we plot $R^{dc}$ in gray scale for values near the AP to P transition, thereby generating an ($I^{dc}$, $f_{inj}$) phase diagram. The diagram shows the reduction trend in $I_{c}$ as $f_{inj}$ increases in the range below $(f_{0}^{dc})^{min}\sim $ 4.2 GHz. This reduction in $I_{c}$ is well above the sweep-to-sweep variation. For this specific sample, we estimate the magnitude of the variations in $I_{c}$ by measuring $\sim$ 80 loops with $f_{inj}$= 4 and 5  GHz.  We account for a slight frequency dependence of the sample impedance by applying at each $f_{inj}$, the power required to pass an equal magnitude $(I^{rf})^{rms}$= 1 mA through the sample.  The difference between the average switching currents $<I_{c}^{\textrm{4 GHz}}>$ and $<I_{c}^{\textrm{5 GHz}}>$ is greater than $15\%$. The standard deviations on the other hand, are below $3\%$.

For $f_{inj}$ in the $f_{0}^{dc}$ range, we identify the plateau, labeled as the \emph{frequency locking region}, as well as an increase in $I_{c}$.  The open squares show the approximate locations of the $f_{0}^{dc}(I^{dc})$ peak resonances, which are centered in the locking region as expected. While the appearance of the plateau in $R^{dc}$ could be attributed to the MR effect resulting from an rf induced modification of the dynamical trajectory of the magnetization, frequency locking between the external rf current and the dc driven precession is also expected to produce a dc rectification voltage \cite{Rippard,Tulapurkar}. Rectification effects due to frequency locking have been studied in similar samples, when appearing in the non-hysteretic regime \cite{Rippard,Tulapurkar,Sankey}.

Further physical insight into the rf driven effects can be gained through MS based on the Landau-Lifshitz-Gilbert equation using Slonzcewski \cite{Slonzcewski} type STT. We compare simulation results for the cases of $I^{rf}$ =0 and $I^{rf}\neq 0$.  For an initial AP configuration and no rf, and using $\alpha$=0.025, we observe switching for $I^{dc}>$ 1.4 mA within a 10 ns window. For lower $I^{dc}$ we find PS dynamics.  The calculated Fourier spectral density of the in-plane component of the magnetization along the short axis, $M_{y}(t)$ drop from 4.1 GHz to 3.1 GHz as the current increases between 1.2 to 1.3 mA. The numerical experiments also provide information concerning the effects of $I^{rf+dc}$ on the dynamics in time and frequency domain.  An important conclusion we extract from the MS is that the precessional frequency and orbit are correlated such that small (large) orbits correspond to high (low) frequencies (which is also the case for dc-only driven PS precession). Therefore, when $I^{rf}$ with $f_{inj} \sim f_{0}^{dc}$ is applied such that \emph{frequency pulling} or \emph{frequency locking} occur, effects on the magnetization precessional orbits also appear.  As the MS do not include the Oersted field from $I^{dc+rf}$, we conclude that all rf induced changes on the magnetization trajectory originate from STT. However, rf \emph{magnetic field} induced reductions in switching fields have also been observed \cite{zhu,thirion}.

In Fig.$\:$4(b), we show the $T=0$ simulated switching boundary by plotting the MR (average value for the final 5 ns of a 20 ns simulation window, normalized to the experimentally obtained MR amplitude) as a function of $I^{dc}$ and $f_{inj}$.  The rf current was set to $(I^{rf})^{rms}=$1 mA to match the experimental amplitude. The light/dark areas correspond to high/low MR states. The switching boundary shows a reduction in $I_{c}$ for $f_{inj} \sim$ 3 GHz, which is close to the numerically obtained $f_{0}^{dc}$, for $I^{dc}$ near $I_{c}$. At slightly higher $f_{inj} \sim$ 3.5 GHz, $I_{c}$ reaches a maximum. A slow decay follows as $f_{inj}$ continues to increase above 3.5 GHz.  The gray area labeled C shows a drop in the MR with respect to the AP state.  When passing from region B to region C, by increasing $f_{inj}$ at a fixed $I^{dc}$, the simulated PS orbits show an an abrupt expansion above $ 10^{\circ}$. This expansion appears as a sharp contrast between regions B and C, revealing the boundary for a major change in the induced precessional orbits. The Fourier transform spectrum of $M_{y}$ shows that the dynamics within the gray area (C) contains a dominant spectral component at $f_{inj}$.
The scattered gray points appearing at $I^{dc}>I_{c}^{dc}$ for $f_{inj} \sim$ 3 GHz, correspond to sustained high amplitude MR oscillations. Additional MS (not shown) reveal that these gray points, together with other gray areas near the switching boundary, become dark (switch) at higher temperatures, thereby accentuating the $I_{c}$ minima occurring between $3$ and $3.5$ GHz. Furthermore, the scattered points observed at T=0 may be a numerical artifact, as their locations depend on the initial conditions. Hence, the reduction for $f_{inj}$ below $f_{0}^{dc}$ and the increase for $f_{inj} \sim f_{0}^{dc}$, in $I_{c}$, as well as the frequency locking boundary are qualitatively reproduced by the simulations.

To understand how the rf affects the switching boundary, recall that as the $I^{dc}$ is ramped toward the switching value, the corresponding resonance frequency drops toward $(f_{0}^{dc})^{min}$. If the injected current additionally has an rf component with frequency $f_{inj}$ just below $(f_{0}^{dc})^{min}$, then frequency pulling will open ($f_{inj}$ pulls toward a lower frequency) the orbit as $I^{dc}$ approaches $I_{c}$.  This pull drives the system toward $f_{inj}$, however, since $f_{inj} < (f_{0}^{dc})^{min}$, there is no stable orbit corresponding to $f_{inj}$, thus frequency locking does not occur and switching at a reduced $I_{c}$ is observed. This effect occurs at lower $I^{dc}$ values for higher $f_{inj}$ in this range and this may be the origin of the consistent decrease in $I_{c}$ obtained as $f_{inj}$ increases in the range below $(f_{0}^{dc})^{min}$.

In contrast, for $f_{inj}\gtrsim (f_{0}^{dc})^{min}$, frequency locking occurs for $I^{dc}$ near the switching boundary, stabilizing the orbit and increasing $I_{c}$. As long as the precessional frequency is locked to $f_{inj}$, the orbit does not expand as $I^{dc}$ increases. Therefore, switching does not occur until $I^{dc}$ is sufficiently high and the corresponding $f_{0}^{dc}$ is far enough from $f_{inj}$ for the system to unlock and the magnetization to switch. For higher frequencies, such that $f_{inj}$ does not remain locked as $I^{dc}$ approaches $I_{c}$, there are no expected effects (of this type) of the rf on $I_{c}$, as is the case for even higher injected frequencies $f_{inj}> (f_{0}^{dc})^{max}$ for which no frequency locking occurs.

In conclusion, we demonstrate that rf currents tuned to the frequency range of the dc driven magnetization precession frequency range, can significantly alter the dc driven PS and switching modes. Based on frequency domain measurements and MS that well describe our experimental data, we explain the origin of the rf driven effects on the precessional orbits and frequencies and on the dc critical switching boundary. This mechanism for reducing $I_{c}$ by additionally injecting rf currents with frequencies slightly below $(f_{0}^{dc})^{min}$ may be of technological interest for the design and control of STT devices.

\begin{acknowledgments}
We acknowledge helpful communications with J. Sun and K. Ito, and thank I. Krivorotov for providing the macrospin simulation code.
\end{acknowledgments}
\bibliography{RFBIB_FLOREZs}
\end{document}